\documentclass[twocolumn]{aastex631}

\graphicspath{{./}{Figures/}}

\submitjournal{ApJL}

\shorttitle{Building an Acceleration Ladder}
\shortauthors{Craig et al.}

\usepackage{graphicx}	
\usepackage{amsmath}	
\usepackage{comment}    

\begin{document}

\title[Building an Acceleration Ladder]{Building an Acceleration Ladder with Tidal Streams and Pulsar Timing}

\correspondingauthor{Peter Craig}
\email{pac4607@rit.edu}

\author{Peter Craig}
\affiliation{Department of Physics and Astronomy, Rochester Institute of Technology, Rochester, NY 14623}
\author{Sukanya Chakrabarti}
\affiliation{Department of Physics and Astronomy, University of Alabama, Huntsville, Huntsville, Alabama 35899}
\author{Robyn E. Sanderson}
\affiliation{Department of Physics and Astronomy, University of Pennsylvania, Philadelphia, PA 19104}
\author{Farnik Nikakhtar}
\affiliation{Department of Physics, Yale University, New Haven, CT 06520}

\begin{abstract}

We analyze stellar streams in action-angle coordinates combined with recent local direct acceleration measurements to provide joint constraints on the potential of our Galaxy. Our stream analysis uses the Kullback-Leibler divergence with a likelihood analysis based on the two-point correlation function. We provide joint constraints from pulsar accelerations and stellar streams for local and global parameters that describe the potential of the Milky Way (MW). Our goal is to build an ``acceleration ladder", where direct acceleration measurements that are currently limited in dynamic range are combined with indirect techniques that can access a much larger volume of the MW. To constrain the MW potential with stellar streams, we consider the Palomar 5, Orphan, Nyx, Helmi and GD1 streams. Of the potential models that we have considered here, the preferred potential for the streams is a two-component Staeckel potential. We also compare the vertical accelerations from stellar streams and pulsar timing, defining a function $f(z) = \alpha_{1pulsar}z - \frac{\partial\Phi}{\partial z}$, where $\Phi$ is the MW potential determined from stellar streams, and $\alpha_{1~\rm pulsar}z$ is the vertical acceleration determined from pulsar timing observations. Our analysis indicates that the Oort limit determined from streams is consistently (regardless of the choice of potential) lower than that determined from pulsar timing observations. The calibration we have derived here may be used to correct the  estimate of the acceleration from stellar streams.

\end{abstract}


\section{Introduction}

Signatures of tidal interactions between galaxies carry valuable information about the Galactic potential and the interaction itself, encoded in the orbital properties of stream member stars \citep{Johnston1999,Newberg2002,Price-Whelan2013,Sanderson2015}, or in disturbances in the stream gas \citep{Levine2006,Weinberg_Blitz2006,Chakrabarti_Blitz2009,Chakrabarti_Blitz2011,Chakrabartietal2011,Chakrabarti2013,Craig2021}. Streams (both stellar and gaseous) have been used to constrain the Milky Way (MW) halo potential out to large distances \citep{Koposov2010,Malhan2019,Reino2021,Vasiliev2021}, and have also been used to constrain dark matter sub-structure \citep{Carlberg2012,Carlberg2013,Sanders2016,Erkal2017,Bonaca2019,Banik2019,Bonaca2020}.

Recently, extreme-precision time-series observations to directly measure Galactic accelerations and the Galactic potential have become feasible. \cite{Chakrabarti2021} analyzed compiled pulsar timing observations to measure the Galactic acceleration and derive fundamental Galactic parameters, including the Oort limit (the mid-plane density), the local dark matter density, and the shape of the Galactic potential. In the near future, we can expect direct acceleration measurements from extreme-precision radial velocity observations \citep{Chakrabarti2020} and from eclipse timing \citep{Chakrabarti2022}. Although direct acceleration measurements have provided local constraints on the potential  \citep{Chakrabarti2021}, they do not yet provide constraints for the MW mass or other global parameters for which we require information over a large volume. While stellar streams provide indirect constraints, they add valuable information about the shape and extent of the MW potential at larger distances. Here, we explore the idea that combining complementary information from local acceleration measurements and tidal streams can be used to build an ``acceleration ladder" to derive constraints on the Galactic potential.

Several methods have been used to extract information about the Galactic potential from tidal streams. One is to fit an orbit to a stream based on the observed positions and velocities along the stream, as done in \cite{Newberg2010} and \cite{Koposov2010}. Agreement between the stream orbit and the data is potential dependent, allowing a fitting procedure to estimate parameters for the MW potential. There are known biases in this methodology that result from differences between the orbits of stream stars and the progenitor \citep{Sanders2013}. An alternative method is to back-integrate the orbits for the stream member stars in the MW potential \citep{PriceWhelan2014}. If these orbits are calculated using an accurate MW potential, then the member stars should become bound to the progenitor at some point in their orbital history. Another common technique is a forward modelling method, such as the one described in \cite{Bonaca2014}, which has several related variations \citep{Varghese2011,Sanders2014,Fardal2015}. This method uses a Monte Carlo Markov Chain algorithm that compares simulated streams to observed streams in 6D phase space.

We focus here on stellar streams rather than gaseous streams to take advantage of the vast infrastructure that has been developed for the collisionless components of galaxies \citep{BinneyTremaine2008}, and the better distance estimates that are available for stars in stellar streams relative to the kinematic distance estimates that are typically available for the gas \citep{Levine2006}. We employ the action space clustering method developed in \cite{Sanderson2015}, and later applied in \cite{Reino2021}. In the true potential, stream stars that were once part of a disrupted dwarf galaxy are expected to be on similar orbits, and thus have similar positions in action space. We apply this by searching for the potentials that lead to the strongest clustering in the actions of stream stars. Since actions are integrals over one period of the orbit, this method is similar to the ``rewinder" algorithm \cite{Price-Whelan2013}. The analysis of stellar streams by \cite{Sanderson2017} employs a ``consensus fit" whereby multiple streams provide information on various regions of the MW potential, allowing the potential model to be valid over larger parts of the galaxy; another advantage of this approach is that the consideration of multiple streams leads to a smaller bias than one stream alone \citep{Bonaca2014} (i.e., due to the disrupted dwarf galaxy being near pericenter, which can bias the results \citep{Reino2022})It has been demonstrated based on the Aquarius simulations that these methods are effective at recovering the present day properties of the host galaxy potential \citep{Sanderson2017}. A limitation of these techniques is that they do not extend well to models for the potential that are not in equilibrium or are time dependent, so these effects cannot be captured. Our goal is to recover the present day properties of the potential, as this is what the direct accelerations probe, so this method works well for our purposes.

One of the challenges to using tidal streams for measuring the Galactic potential is obtaining 6D phase space information for the member stars. High quality positions and proper motions can typically be obtained from Gaia data \citep{GaiaMission,DR2Release}, significantly expanding the sample of stars with available information \citep{Bonaca2014}. To get to full 6D information we also require distances and radial velocities. For some streams, missing measurements can be estimated by fitting the distance or radial velocity along the stream track, then interpolating based on the position in the stream. The same method is used in \cite{Reino2021} for the streams considered here, and for Sagittarius in \cite{Vasiliev2021}.

One of our goals is to use recent direct acceleration measurements from pulsar timing to calibrate indirect measurements for the potential of the MW. This would in effect serve to build an ``acceleration ladder", similar to the distance ladder that is based on a calibration of indirect distance measures to the basic geometric measure of distance from parallax measurements. Direct acceleration measurements provide the first rung of the ladder or a ``pivot point" for the Galactic potential, which can be used to calibrate stream-based potential measurements. In \cite{Reino2021}, only models that matched the circular velocity at the Solar circle to within uncertainties were considered, but the measurements of the circular velocity are indirect, while our pulsar accelerations provide a direct measurement of the potential. The paper is organized as follows. We present our methodology in \S 2 and discuss our data selection in \S 3. We present our main results in \S 4, and conclude in \S 5. 

\section{Methods}

We use the python package \textsc{galpy} \citep{galpy} to calculate the actions for our streams. They are calculated analytically for the Isochrone potential \citep{BinneyTremaine2008}, and numerically in all other potentials considered. In spherically symmetric potentials, like the Hernquist potential, we use \textsc{galpy's} spherical approximation method, providing accurate and efficient action calculations. Staeckel potentials allow us to use the Staeckel approximations, providing accurate actions requiring only a single numerical integral for each action \citep{Bonaca2014}. For all other potentials, we use the Isochrone approximation method in \textsc{galpy}. This method uses an auxiliary Isochrone potential with integrated orbits of the stars in the desired potential to compute the actions. This requires orbit integration for each star, and is therefore relatively computationally intensive. However, this provides more accurate approximations to the actions in these potentials compared to other methods. In particular, the assumptions made for the other methods break down when the radial or vertical actions are on the same order as L$_{z}$ \citep{Bovy2014}. 

For each potential model, we tested the accuracy of the action approximations by calculating the actions of all our stream members along their orbits for a variety of different points in the parameter space. We use a leap-frog algorithm to compute the orbit for at least 1 orbital period, and then calculate the actions at regular intervals along the orbit. The actions should remain constant, and we find that the action variations are consistently less than $ 1 \%$ in J$_{R}$ and J$_{Z}$. When we use data from multiple streams, we shift the actions in $L_{z}$ for each individual stream such that there is limited overlap between streams. This will not impact the clustering in action space, but helps to avoid error modes with large masses and low scale lengths \citep{Reino2021}. 

As we intend to build upon the results from direct acceleration measurements from pulsar timing, we consider the potentials used in \cite{Chakrabarti2021}. Specifically, we consider the $\alpha_1$ potential and the $\alpha\gamma$ potential. These are defined in Equations \ref{eqn:alpha1} and \ref{eqn:alphagamma} respectively, where $\alpha_{1}$ is the inverse square of the frequency of low-amplitude vertical oscillations, $V_{\rm LSR}$ is the local standard of rest velocity, and $\gamma$ describes the shape of the potential. The $\alpha_1$ potential provided the best fit to the pulsar accelerations, and the $\alpha\gamma$ potential is the ``cross-term'' model discussed in \cite{Chakrabarti2021}. These models provide a good match to the measured pulsar accelerations, however they do not produce strong clustering in action space. This is expected because these models were designed to produce reasonable orbits for stars in the Solar neighborhood with low eccentricities and inclinations, which is not generally satisfied by our stream members \citep{Quillen2020}.
\bigskip

\begin{equation}\label{eqn:alpha1}
\Phi(R,z) = V_{\text{LSR}}^{2} \log(R/R_{\odot}) + \frac{1}{2} \alpha_1 z^2
\end{equation}

\begin{equation}\label{eqn:alphagamma}
\Phi(R,z) = V_{\text{LSR}}^{2} \log(R/R_{\odot}) + \log(R/R_{\odot}) \gamma z^2 + \frac{1}{2} \alpha_1 z^2
\end{equation}

\subsection{Kullback-Leibler Divergence}

We use the Kullback-Leibler Divergence (KLD) to measure the amount of clustering in action space, as in \citep{Sanderson2015}. The KLD compares two PDFs to each other. We use it to measure the amount of action space clustering by comparing the action distribution with a uniform distribution over the maximum range for each action across all the datasets (potentials and streams). Larger KLD values indicate a larger difference between the distributions, implying that there is more clustering in action space. Equation \ref{eqn:KLD} defines the KLD for a continuous random variable from p(x) to q(x), where in our case p(x) will be the action distribution and q(x) will be our uniform distribution.

\begin{equation}\label{eqn:KLD}
KLD(p:q) = \int p(x)log(\frac{p(x)}{q(x)})dx
\end{equation}

We use the kernel density estimator \textsc{enbid} \citep{Sharma2011} to estimate the PDF for the actions. We define the resulting distribution for a potential with parameters a as $f_a(J)$, and then we can calculate the KLD from equation \ref{eqn:KLD1}. This integral can be re-written as in \citep{Reino2021} using a Monte Carlo approximation to sum across points drawn from the distribution. This is a natural choice in our case where we have a discreet set of actions corresponding to our member stars. The exact form of the calculation used corresponds to the wKLD1 given in \cite{Reino2021}, which includes a weight term such that the streams contribute equally in joint fits. 

\begin{equation}\label{eqn:KLD1}
D^{1}_{KL} = \int f_a(J)log(\frac{f_a(\bf{J})}{f_a^{\emph{shuf}}(\bf{J})})d^{3}J
\end{equation}

For our data set, we use an optimization algorithm to find the maximum value of the KLD within reasonable bounds. For 2 parameter potentials, we apply a simple grid search across the parameter space. The same grid can be used for the error analysis technique described in \cite{Reino2021} and \cite{Sanderson2015}, which measures the KLD between the best fitting action distribution and the distributions in other potentials. For potentials with more than 2 parameters, we use a differential evolution optimizer implemented in \textsc{scipy} \citep{Scipy} to maximize the KLD.

\subsection{Likelihoods and Error Analysis}

We use the two-point correlation function to compute the likelihood from the action distributions in order to compare with the pulsar sample. We use the likelihoods from both methods and combine them to get a joint likelihood function. The likelihoods for the streams are calculated using the methods given in \citep{Yang2020}. This approach is different from that used in \cite{Reino2021}, chosen here because it enables the combination of our methods. 

We can use the likelihoods to estimate the uncertainty by examining the relative likelihood (i.e., the likelihood divided by the maximum likelihood value) for the parameters that we are interested in. A confidence region in that parameter space can be defined by setting a threshold on the relative likelihoods corresponding to the desired confidence interval. We generally focus on the total mass and Oort limit for the various potentials, but also compute a surface across all parameters to measure their uncertainties. We considered using Fischer matrix analysis to compute the uncertainties. However, this analysis underestimates the uncertainties for the streams. The likelihood surfaces can have multiple local maxima near the maximum likelihood, corresponding to the best fits of individual streams. Since the Fisher matrix analysis is based on derivatives of the likelihood function, it is only sensitive to regions near the maximum likelihood, leading to an underestimation of the uncertainties.

The combined sample (i.e., for pulsars and streams) has a likelihood surface defined by the product of the two individual likelihoods, which gives us the relative likelihoods for the combined sample. In this way we can easily obtain the uncertainties for both the individual and combined samples using the same mechanism.

For all the potentials considered except for the $\alpha_1$ potential, we examine the total mass and Oort limit for the models. In cases with more than 2 parameters we marginalize over the remaining parameters to produce a two-dimensional likelihood surface. We expect the streams to produce better constraints on the total mass, while the pulsars should provide stronger constraints on the Oort limit. Thus, we expect the combined sample to provide tighter constraints than either individual sample.

Note that we do not use a maximum likelihood analysis to find the best fitting parameters, instead maximizing the KLD as described above. From our likelihood surfaces we have confirmed that the best fitting KLD parameters closely agree with the maximum value of our likelihood surface, so we retain the best-fit identified using the method of \citep{Reino2021}. 

The equations from \cite{Yang2020} used to calculate the likelihood are given in Equations \ref{eqn:plnd} and \ref{eqn:logl}. Here D is a distance between particles in action space normalized by the standard deviations of all the actions. $\mathcal{P}(\ln D)$ is a probability distribution describing the distance between stars in action space, which is computed using equation \ref{eqn:plnd} and the two-point correlation function. We set a maximum value of $D_{max}$, as at large distances the behavior will change due to an insufficient number of pairs.

\begin{equation}\label{eqn:plnd}
1 + \xi{\ln D} = \frac{D_{\text{max}}\mathcal{P}(\ln D)}{3D^3\int^{\ln D_{\text{max}}}_{-\infty} \mathcal{P}(\ln{D\prime})d \ln D\prime}
\end{equation}

\begin{equation}\label{eqn:logl}
\ln(\mathcal{L}) = N_{\text{pairs}} \int^{\ln D_{\text{max}}}_{-\infty} \mathcal{P}(\ln D) \ln[ 1 + \xi(\ln D)] d \ln \text{D}
\end{equation}

\begin{figure}
\begin{center}
\includegraphics[width=\columnwidth]{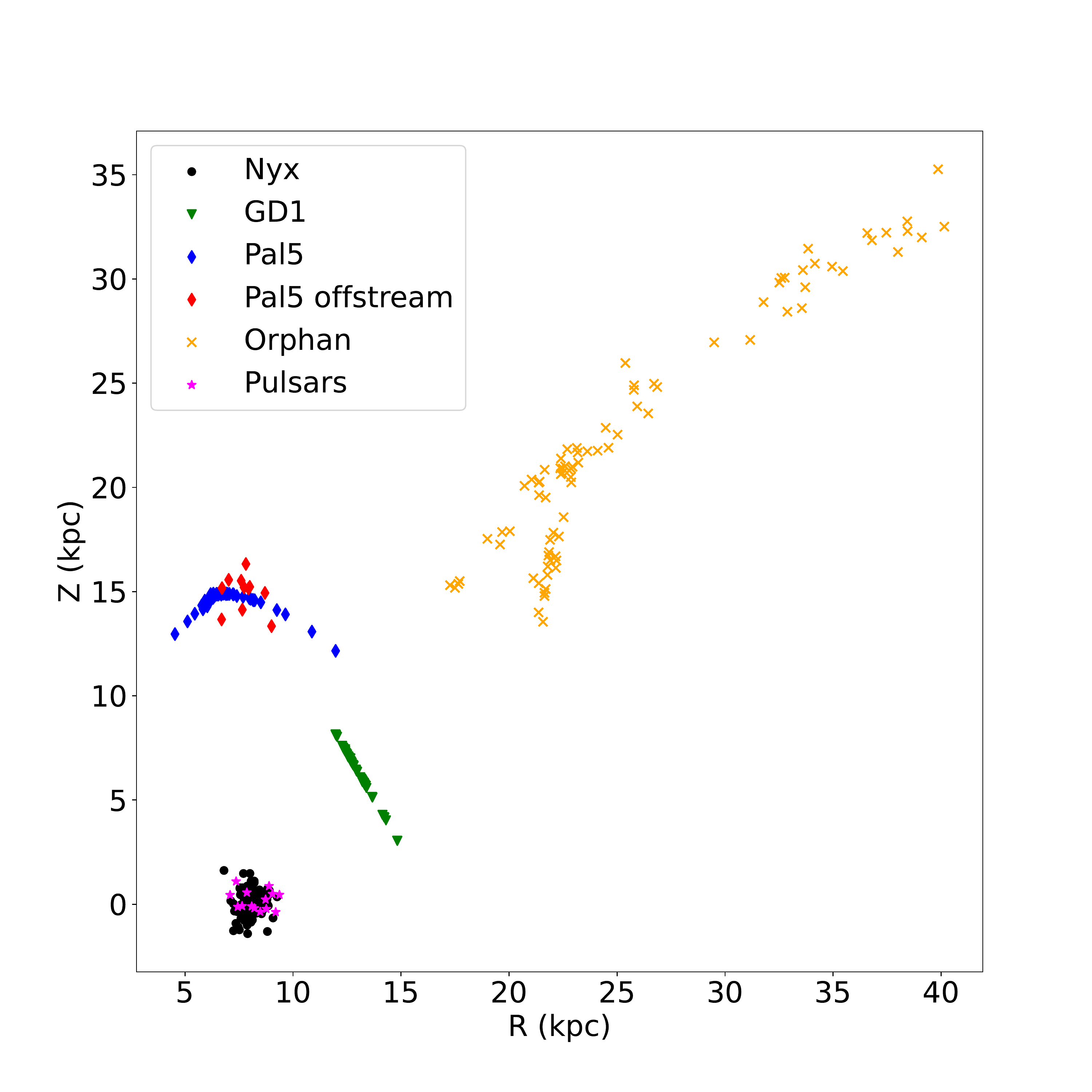}
\caption{Stream and pulsar samples shown in Galactocentric coordinates. This is the main data set used for the results presented in this paper. The red diamonds show members of the Pal5 stream that are somewhat separated from the clear stream track in this space.} \label{fig:pulsar_overlap}
\end{center}
\end{figure}

\section{Data Selection}

We have considered five different tidal streams - GD1, Orphan, Palomar 5, Nyx \citep{Necib2020} and the Helmi stream \citep{Helmi1999}. We find that for some potentials the Helmi stream produces large uncertainties, and is not included in our primary results. Recent work finds that the Helmi stream provides a constraint for the shape of the inner dark matter halo of the Milky Way \citep{Dodd2022}, noting that this stream may be in a resonance. However, if it is truly in a resonance, calculating actions for typical potentials is not valid, which is consistent with our findings here. Meanwhile, \cite{Zucker2021} find that the abundances of likely Nyx stream members are consistent with the thick disk, and may not have an extragalactic origin. It does, however, remain plausible that this is the result of an early minor dwarf galaxy merge \citep{Nyx2}. Therefore our results are calculated using the Nyx, GD1, Palomar 5 and Orphan Streams. However, to avoid contamination if the Nyx stream is actually not extragalactic in origin, all calculations are also performed without these stars included. The rest of our data set is similar to that in \citep{Reino2021}, to reproduce the stream sample and analysis in \cite{Reino2021} as closely as possible. We have produced results in our primary potential model that includes the Nyx stream as well. In this potential, the Nyx Stream does not make a large impact on the derived potential parameters, but does allow for phase space overlap between samples. This overlap allows for better calibrations between the streams and the pulsar accelerations. The Galactocentric R and Z values of the stream member stars can be seen in Figure \ref{fig:pulsar_overlap}. 

The Pal5 stream here is split into two separate samples, a main stream group and a set of 11 stars that appear to be slightly off of the main stream track. To check for any potential biases induced by the inclusion of these sources, we have tracked them in the action calculations, and found that they are not more likely to be outliers in action space than the other stars. This can be seen from the action distributions shown in Figure \ref{fig:actions}. We have also considered the two separate branches that appear in the Orphan stream data set. All of the analysis that follows has been performed using each branch separately in addition to the full sample. Changing the sample does slightly alter the resulting potential parameters, but at a much smaller level than the estimated uncertainties. These changes do not impact the main results of this paper, so we have retained the full sample here.

\section{Results}

\begin{figure}
\begin{center}
\includegraphics[width=\columnwidth]{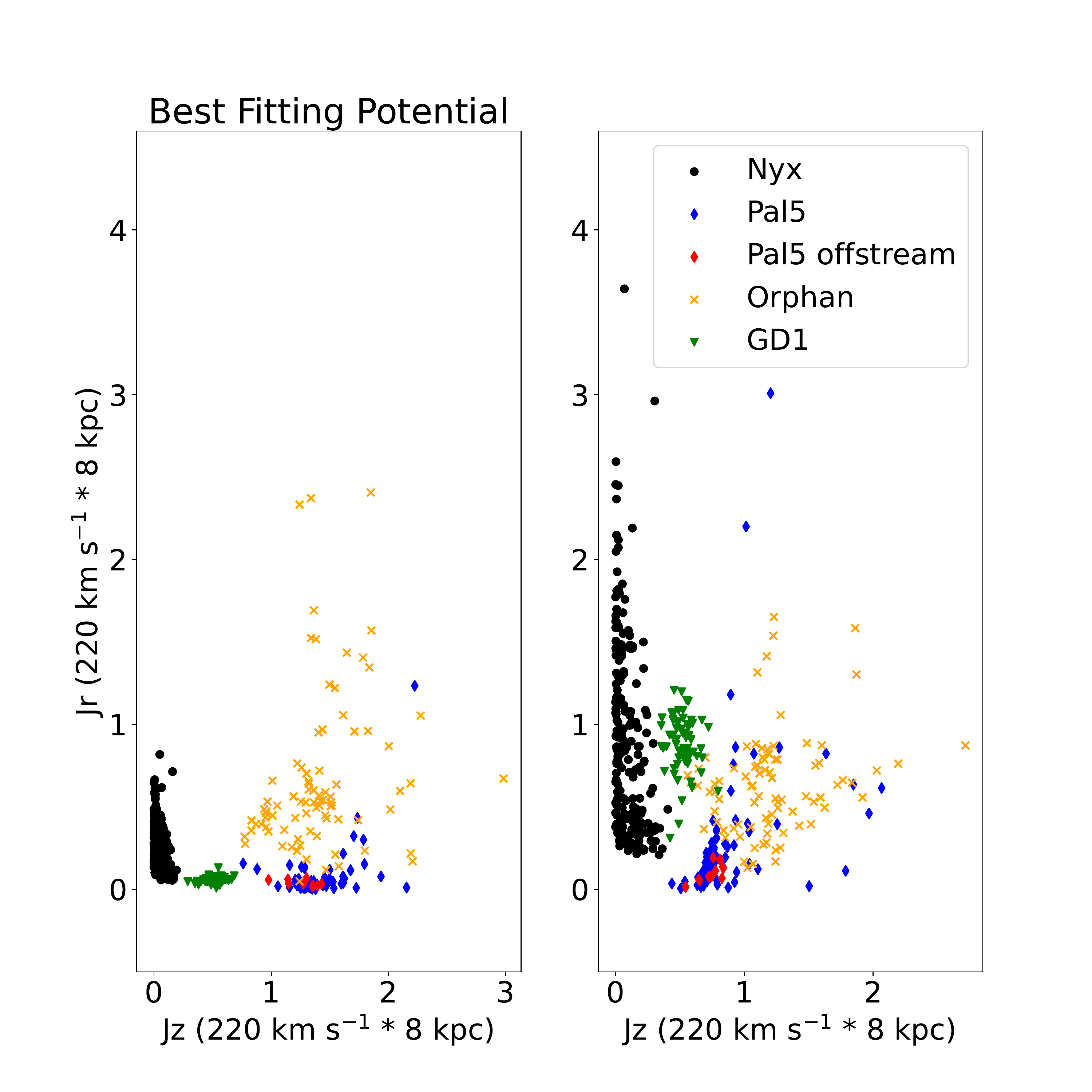}
\caption{Vertical and radial actions for stars in several streams, computed using a two-component Staeckel potential. The left panel uses the best fitting potential for the streams, with parameters listed in Table \ref{tab:stats}. The right panel shows a distribution that differs from the best-fit potential by roughly $2 \sigma$, based on our error analysis. As before, the red diamonds indicate the locations of the members of Pal5 that appear to be off the main stream. Interestingly, these sources are not the main outliers from the central Pal5 clump in action space.} \label{fig:actions}
\end{center}
\end{figure}

\begin{figure}
\begin{center}
\includegraphics[width=\columnwidth]{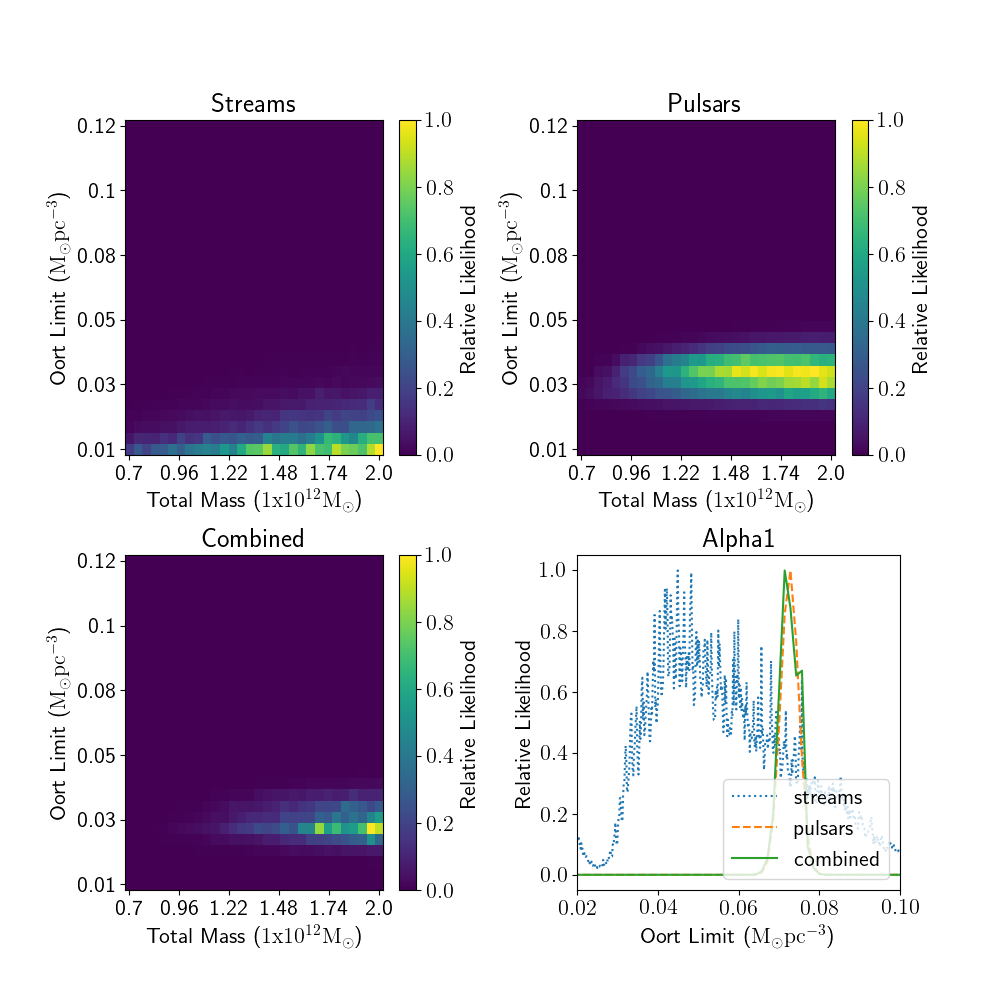}
\caption{Likelihood surfaces for the two-component Staeckel and $\alpha_{1}$ potentials. Included are the stream, pulsar, and combined relative likelihood surfaces in both potentials. In this case we can see the general trends in the Oort limit that appear across all of our models, where the streams have a lower value for the Oort limit along with higher uncertainties.}\label{fig:ellipse}
\end{center}
\end{figure}

\begin{table*}
\centering
\caption{Best-Fit Parameters and KLD Values}\label{tab:stats}
\begin{tabular}{c c c c}

\hline
\hline
Potential & KLD & Best-Fit Parameters & $\rho_{\text{local}}$ (M$_{\odot}$ / pc$^3$) \\
\hline
Hernquist & 0.95 & M = 7.0 $\times 10^{11}$ M$_\odot$ & $0.021 \pm 0.02$ \\
& & a = 8.66 kpc &\\
$\alpha\gamma$ & 0.22 & $\log_{10}(\alpha_1/\text{Gyr}^{-2}) = 3.32 $ & $0.037 \pm 0.023$\\
& & $\log_{10}(\gamma/\text{Gyr}^{-2}) = -5.52$ &\\
$\alpha\beta$ & 0.221 & $\log_{10}(\alpha_1/\text{Gyr}^{-2}) = 3.31 $ & $0.036 \pm 0.03$\\
& & $\beta = -0.16$ &\\
$\alpha_1$ & 0.224 & $\log_{10}(\alpha_1/\text{Gyr}^{-2}) = 3.36$ & $0.045 \pm 0.015$\\
Isochrone& 1.07 & M = 1.55 $\times 10^{12}$ M$_\odot$ & $0.029 \pm 0.02$\\
& & b = 11.92 kpc &\\
Two-Component Staeckel & 1.32 & M$_t$ = 2.23 $\times 10^{12}$ M$_\odot$ & $0.026 \pm 0.02$\\
& & k = 0.33 &\\
& & a$_{\mathrm{inner}}$ = 18.00 kpc &\\
& & a$_{\mathrm{outer}}$ = 37.78 kpc &\\
& & e$_\mathrm{inner}$ = 1.95 &\\
MWPotential2014& 0.53 & \citet{galpy} Values & 0.101\\
Best-Fit Pulsars & -- & $\log_{10}(\alpha_1/\text{Gyr}^{-2}) = 3.61$ & $0.08^{+0.05}_{-0.02}$\\
\hline
\end{tabular}\par
\bigskip
\textbf{Notes.} KLD optimization results for a number of different potentials from the GD1, Palomar 5 and Orphan streams. A two-component Staeckel model provides most clustered actions with a KLD of 1.51. Most of these results lack the Nyx stream in order to minimize potential contamination. The two-component Staeckel model includes this data set however, as it is our best fit stream result without the Nyx stream, and the results are only slightly altered as a result of the inclusion of this data set. This allows for potential constraints with phase space overlap with the pulsar sample.
\end{table*}

Our main results are based on a combination of multiple streams. A single stream will often produce unrealistic potential parameters, while the constraints derived from a combination of streams generates a much more realistic potential \citep{Bonaca2014,Bonaca2018,Reino2022}. We have calculated the best fitting potential for each of our streams individually, as well as with a combination of streams. This is repeated across a range of different potential models, as seen in Table \ref{tab:stats}. From these results we consider the potential model giving the largest KLD value to be our best fitting potential, in this case a two-component Staeckel potential. This can then be compared with the properties of the potentials that were found to be a good fit with the pulsar sample. Shown in Figure \ref{fig:actions} are the radial and azimuthal actions of the Palomar 5, GD1, Nyx and Orphan streams in a two-component Staeckel potential, in both the best-fit potential and a poorly fitting potential.

The results for our combined sample can be seen in Figure \ref{fig:ellipse}, which shows the likelihood surface for the Oort limit and MW mass using streams and pulsar timing. We consider both the $\alpha_{1}$ potential, which is the best fit to the pulsars, and the two-component Staeckel potential, which is the preferred potential model for the streams. 

\begin{figure}[ht]
\begin{center}
\includegraphics[width=\columnwidth]{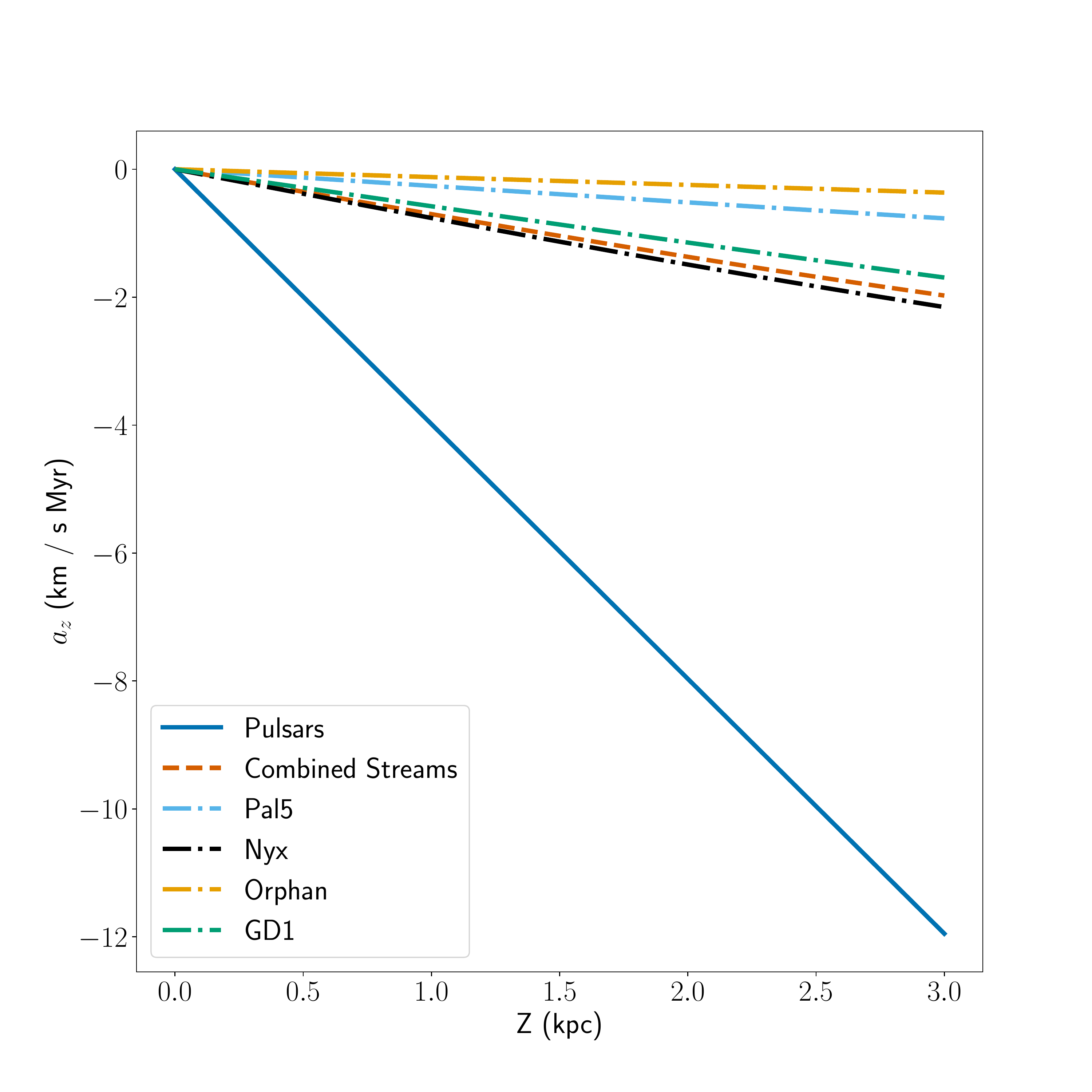}
\caption{Vertical accelerations for the pulsars and the streams. The pulsars are shown in their best fitting $\alpha_1$ potential, while the streams are shown in their best fitting two-component Staeckel potential. We show the sample with the combined stream data set as well as each stream individually.} \label{fig:fz}
\end{center}
\end{figure}

Interestingly, the streams lead to a lower Oort limit than the pulsars in every models considered. This is not necessarily surprising as the streams probe distances out to tens of kpc, while the Oort limit is sensitive to the scale height of the disk. Additionally, Staeckel models do not typically produce a thin disk as the measured pulsar accelerations indicate \citep{Batsleer1994}. They are capable of producing models with Oort limits that match the pulsar value however. The pulsars primarily constrain the acceleration in the vertical direction, so we compare the accelerations from the two methods in the vertical direction. We can write the difference in the accelerations as the function specified in Equation \ref{eqn:f(z)}. A plot for the vertical accelerations of our preferred models can be seen in Figure  \ref{fig:fz}, where $\alpha_{\mathrm{1,pulsar}} = 4073^{+1422}_{-837} Gyr^{-2}$ is the value of $\alpha_1$ that is derived from measured pulsar accelerations, and $\Phi$ is the best-fit potential for the streams. Consistent with the lower Oort limit values, the stream samples yield smaller vertical accelerations.

\begin{equation}\label{eqn:f(z)}
f(z) = \alpha_{\mathrm{1,pulsar}}z - \frac{\partial\Phi}{\partial z}
\end{equation}

\section{Conclusions}

$\bullet$ We analyze clustering in action space to determine the best-fit potential for a set of stellar streams that have also been analyzed by \cite{Reino2021}, with the addition of the Nyx stream. We compare the derived fundamental parameters that describe our Galaxy from stellar streams to recent direct acceleration measurements from pulsar timing \citep{Chakrabarti2021}.

$\bullet$ We consider several potential models, including Hernquist potentials, Isochrone potentials, Staeckel potentials and the potentials used in analyzing the pulsar sample. We analyze Staeckel potentials with both one and two components. The preferred potential for the streams is a two-component Staeckel model, which is the same model considered in \cite{Reino2021} and produces the most clustered actions for our streams.

$\bullet$ We focus primarily here on the mass of the Galaxy and the Oort limit, where the streams provide stronger constraints on the mass and the pulsars provide optimal constraints on the Oort limit. We find that the best-fit potentials for stellar streams  underestimate the Oort limit compared to the value derived from measured accelerations from pulsar timing. For the potential masses in the two-component Staeckel model, we obtain $\mathrm{M}_\mathrm{streams} = 2.15 \pm 0.75 \times 10^{12} \mathrm{M}_{\odot}$, $\mathrm{M}_\mathrm{pulsars} = 1.95 \pm 1.3 \times 10^{12} \mathrm{M}_{\odot}$ and $\mathrm{M}_\mathrm{combined} = 1.98 \pm 0.65 \times 10^{12} \mathrm{M}_{\odot}$. Our estimated Oort limits in this potential are $\rho_{\mathrm{streams}} = 0.022 \pm 0.02 \mathrm{M}_{\odot} \mathrm{pc}^{-3}$ and  $\rho_{\mathrm{pulsars}} = 0.036 \pm 0.011 \mathrm{M}_{\odot} \mathrm{pc}^{-3}$. These values can be compared to the Oort limit of the pulsar's preferred potential determined from direct acceleration measurements, which is $\rho_{\mathrm{combined}} = 0.08^{+0.05}_{-0.02} \mathrm{M}_{\odot} \mathrm{pc}^{-3}$. The pulsar sample provides better constraints since these local accelerations are highly dependent on the local density, which drives the vertical component of the acceleration.

$\bullet$ The Oort limit obtained from the joint constraints of streams and pulsars is lower than the Oort limit derived from  pulsars alone, with a value of $0.028 \pm 0.008 \mathrm{M}_{\odot} \mathrm{pc}^{-3}$. However, pulsars alone do not currently constrain the mass of the potential, since the pulsars are distributed within $\sim$ 1 kpc of the Sun.

$\bullet$ We provide a fitting formula that may be used to calibrate vertical accelerations of stellar streams to the measured pulsar accelerations.

\bigskip
\smallskip
\section*{Acknowledgements}
We are grateful to Stella Reino for sharing her scripts for calculating KLD values. SC and RES gratefully acknowledge support from NSF grant AST-2007232. SC also acknowledges support from RCSA's  Scialog Time Domain Astrophysics Program.
RES acknowledges support from NASA grant 19-ATP19-0068, from the Research Corporation through the Scialog Fellows program on Time Domain Astronomy, and from HST-AR-15809 from the Space Telescope Science Institute (STScI), which is operated by AURA, Inc., under NASA contract NAS5-26555. This work has made use of data from the European Space Agency (ESA) mission Gaia (https://www.cosmos.esa.int/gaia), processed by the Gaia Data Processing and Analysis Consortium (DPAC, https://www.cosmos.esa.int/web/gaia/dpac/consortium). Funding for the DPAC has been provided by national institutions, in particular the institutions participating in the Gaia Multilateral Agreement.

\bibliography{streams}
\bibliographystyle{aasjournal}

\end{document}